\RequirePackage{fix-cm}
\documentclass[twocolumn,epjc3]{svjour3} 
\smartqed  
\usepackage{graphicx}
\usepackage{dcolumn}
\usepackage{bm}
\usepackage{amsmath}
\usepackage[usenames,dvipsnames,svgnames,table]{xcolor}
\usepackage{enumerate}
\usepackage{mathtools}
\usepackage[colorlinks=true,
            linkcolor=blue,
            urlcolor=blue,
            citecolor=blue]{hyperref}

\newcommand{\be}[1]{\begin{equation}\label{eq:#1}}
\newcommand{\ee}{\end{equation}}
\newcommand{\bea}{\begin{eqnarray}}
\newcommand{\eea}{\end{eqnarray}}

\newcommand{\phd}{\phantom{\dag}}

\newcommand{\up}{^{\phd}}
\newcommand{\noi}{\noindent}
\newcommand{\no}{\nonumber}

\RequirePackage{graphicx}
\begin{document}

\title{Platform for controllable Majorana zero modes using superconductor/ferromagnet heterostructures
}


\author{Giorgos Livanas\thanksref{addr1}
        \and
        Nikolaos Vanas\thanksref{addr1} \and Manfred Sigrist\thanksref{addr2} \and Georgios Varelogiannis\thanksref{addr1} 
}



\institute{Department of Applied Mathematical and Physical Sciences, NTUA, 15780, Athens, Greece \label{addr1}
\and
Institut für Theoretische Physik, ETH, 8093, Zürich, Switzerland \label{addr2}
}


\abstractdc{We propose a novel platform for the creation and manipulation of Majorana zero modes consisting of a ferromagnetic metallic wire placed between conventional superconductors which are in proximity to ferromagnetic insulators. Our device relies on the interplay of applied supercurrents and exchange fields emerging from the ferromagnetic insulators.
We assert that the proposed superconductor/ferromagnet heterostructures exhibit enhanced controllability, since topological superconductivity can be tuned apart from gate voltages applied on the ferromagnetic wire, also by manipulating the applied supercurrents and/or the magnetisation of the ferromagnetic insulators.}

\maketitle
%
%
\section{Introduction}
\label{intro}
\begin{figure}[h!]
\includegraphics[scale=1.0]{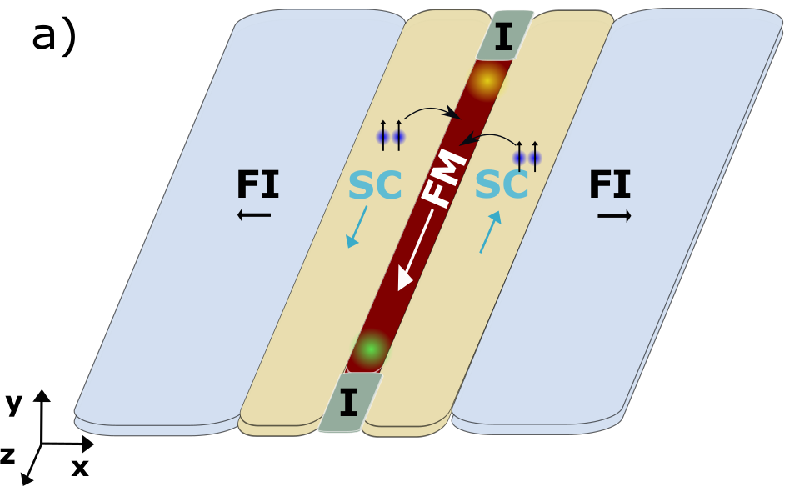}
\includegraphics[scale=0.3]{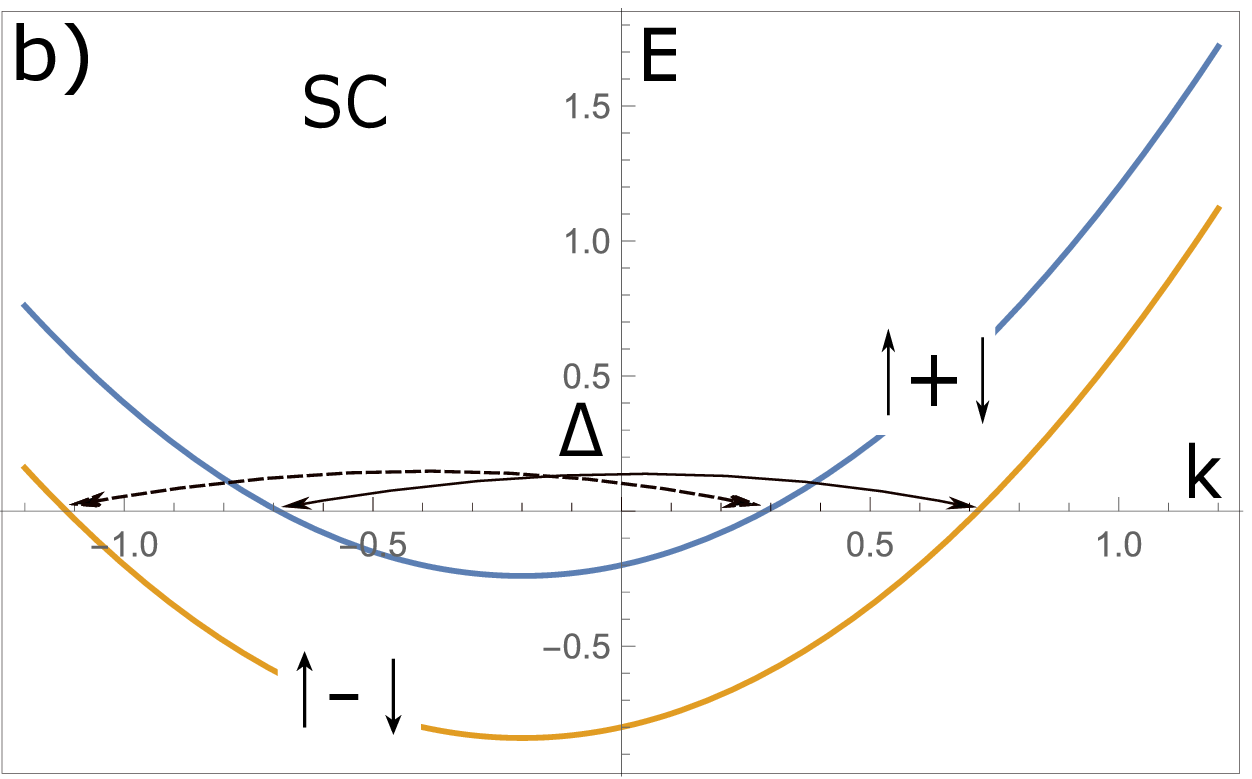}
\includegraphics[scale=0.3]{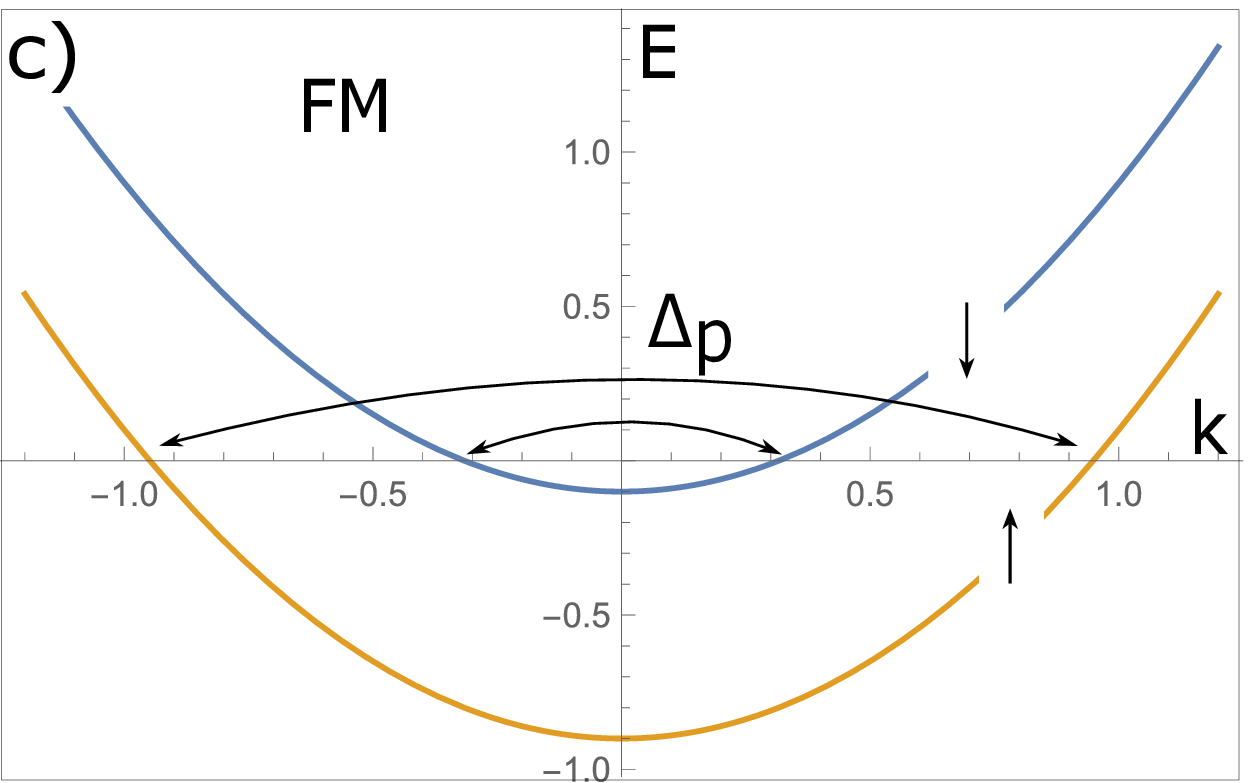}
\caption{ a) Device for engineering controllable MZMs based on superconductor-ferromagnet heterostructures. A ferromagnetic metallic wire (FM) is placed between two conventional superconductors (SCs) where supercurrents (blue arrows) are applied in an anti-parallel configuration. Exchange fields are induced in the SCs due to proximity with the ferromagnetic insulators (FIs). The optimal configuration is when the magnetisations of the FIs (black arrows) is perpendicular to that of the FM (white arrow) and anti-parallel with each other. The application of supercurrent $J$ and exchange field $h_x$ in the SC induces an equal spin triplet pairing component (blue dots) which mediates in the FM resulting in the emergence of MZMs $\gamma_{a}=(\psi^{\dag}+\psi)$ (green dot) and $\gamma_{b}=i(\psi^{\dag}-\psi)$ (yellow dot) at the edges of the wire.  b) The energy bands of a system with an applied current $J$ and a $h_x$ exchange field. The singlet superconducting field $\Delta$ pairs electron from different bands. The application of $J$ and $h_x$ results in an imbalance of singlet pairing between the two spin configurations, and therefore an equal spin triplet pairing component is induced. However, the induced triplet superconducting field still pairs electrons from different bands. c) The induced equal spin triplet superconducting correlations $\Delta_p$ mediate in the FM where an exchange field $h_z$ is present. In the spin configuration of the energy bands of FM, $\Delta_p$ pairs electrons of the same band, i. e. intraband pairing, realising an effectively spinless superconducting state.}
 \label{fig:1}
\end{figure}

Topological quantum computation is the most promising answer to the quantum decoherence problem hindering the development of quantum computers \cite{Nayak}. Topologically protected qubits formed by non-Abelian anyons would allow to store and process quantum information in decoherence-resilient schemes. Majorana zero modes (MZMs) are arguably the most natural realization of such non-Abelian anyons. Indeed, MZMs are expected to emerge at the edges of 1D \cite{Kitaev} and at the vortices of 2D \cite{Ivanov} topological p-wave superconductors but unfortunately p-wave superconductivity appears to be scarce in nature. 
However, a topological p-wave superconducting state has been demonstrated to emerge in heterostructures comprising topological insulators and conventional superconductors \cite{Kane}, where proximity induced singlet superconductivity pairs the helical states confined at the boundaries of the topological insulator realising an effectively spinless superconducting state.
Following this seminal work numerous proposals have been put forward involving superconductor-semicondu \\
ctor heterostructures \cite{SauA,Alicea,Linder}, non-centrosymmetric superconductors \cite{Ghosh}, surfaces of heavy metals \cite{Potter}, the quantum anomalous Hall state \cite{Qi} and 2D electron gases \cite{Hell}.
 Key elements for realising effectively spinless superconductivity in all the aforementioned proposals, is spin momentum locking and a Zeeman energy term, thereby many schemes for realising MZMs rely on effective spin orbit coupling emerging from the application of laser beams in ultracold atoms superfluids \cite{Zhu}, nanomagnets in superconducting wires \cite{Kjaergaard} or carbon nanoribbons \cite{Loss}, magnetic tunnel junctions in proximized 2D electron gases \cite{Matos}, from the helical ordering of magnetic adatoms embedded in conventional superconductors \cite{Choy,Pientka,Stano,Vazifeh,Schecter}
as well as from the curved geometry of carbon nanotubes \cite{SauB} and semiconducting wires \cite{Ojanen}. Moreover, in most of the above setups for engineering MZMs, the required Zeeman energy is provided by the application of an external magnetic field.
Superconductor/ferromagnet heterostructures (SC-FM) support the emergence of topological superconductivity  without the requirement of an external magnetic field, as has been already demonstrated for ferromagnetic metal/superconductor heterostructures with spin active interfaces \cite{Zhang,Takei} and ferromagnetically aligned magnetic adatoms chains embedded in conventional superconductors with spin orbit coupling \cite{Perge}. However, no related devices exhibiting
enhanced controllability of the MZMs, necessary for the future design of braiding operations, has been proposed so far.\\
\indent Here we propose such a platform for engineering controllable TSC and MZMs based on superconductor/ferromagnet heterostructures (SC-FM) and application of supercurrents and gate voltages.  The heterostructures consist of a ferromagnetic metallic wire (FM) placed among conventional superconductors (SCs) which are in proximity to ferromagnetic insulators (FIs) \\ (Fig. \ref{fig:1}). The general underlying mechanism has
already been presented in our previous publication \cite{Livanas} and consists of two stages. In the first stage, triplet p-wave correlations are induced in the SCs, due to the coexistence of conventional superconductivity with applied supercurrents and exchange fields emerging from the FIs. As we show below, the d-vector \cite{Ueda} of the induced p-wave correlations is parallel to polarization of the exchange field of the FIs. In the second stage, these p-wave correlations mediate into the FM where an analogous to Kitaev's chain spinless superconducting state is realised provided the magnetisation of the wire and the d-vector of the p-wave correlations are misaligned. We demonstrate that our device platform exhibits enhanced controllability
through the combination of externally applied supercurrents and the manipulation of the
exchange fields resulting from the FIs that can be achieved by the use of currents \cite{Avci} or applied gate voltages \cite{Chiba,Chiba2}.
We remark that although supercurrents were involved in other proposals for engineering MZMs \cite{Romito,Heimes,Dmytruk,Lesser,Rontynen,Melo}, in our device they do not constitute simply an alternative to magnetic fields, but instead they can also be exploited to manipulate the phase of the induced TSC, a property crucial for designing feasible braiding schemes. 
\section{Induced p-wave superconductivity}
\label{sec:1}
Initially, we describe how the coexistence of conventional superconductivity,
supercurrents and exchange fields induces triplet p-wave correlations
with a d-vector parallel to the polarization of the exchange fields.
As we have
already pointed out in
\cite{Livanas}
 these four fields or order parameters constitute a quartet \cite{Quartets} meaning that
 the presence of any three of them induces the missing fourth one. To this end we employ the following Hamiltonian representing a 1D SC,

\bea
\quad  H=\int \Psi^{\dag} \left( [\frac{\partial_x^2}{2m} \right. +\mu]\tau_z
 + \tau_z \bm{h} \cdot \tilde{\bm{\sigma}} +\Delta\tau_y\sigma_y  \biggl ) \Psi \, \label{Eq:1}
\eea

\noi where $\Psi^{\dag}=(\psi_{\uparrow}^{\dag},\psi_{\downarrow}^{\dag},\psi_{\uparrow},\psi_{\downarrow})$ the extended Nambu spinor and $\bm{\tau}$ and $\tilde{\bm{\sigma}}=(\sigma_x,\tau_z\sigma_y,\sigma_z)$ are the Pauli matrices acting on particle-hole and spin space, respectively, $\bm{h}$ the exchange field and $\Delta=|\Delta|e^{iJx\tau_z}$ the singlet superconducting field.  Under the gauge transformation $\Psi(x) \rightarrow e^{-iJ/2x\tau_z}\Psi(x)$ kinetic term $\frac{\partial_x^2}{2m}$ transforms to $\frac{\partial_x^2}{2m}+i\frac{J}{2m}\partial_x - \frac{J^2}{8m}\tau_z$ where $i\frac{J}{2m}\partial_x$ is a current term and $- \frac{J^2}{8m}\tau_z$ can be absorbed in the chemical potential $\mu$. Thus, Hamiltonian Eq. \ref{Eq:1} takes the form $ H=\int dx \Psi^{\dag} \biggl( [\frac{\partial_x^2}{2m}-\mu]\tau_z  +i\frac{J}{2m}\partial_x + \tau_z \bm{h} \cdot \tilde{\bm{\sigma}} +|\Delta| \tau_y\sigma_y  \biggl ) \Psi $. For periodic boundary conditions, a Fourier transformation to momentum $k$ space  leads to the energy bands of the wire $E_{s,\pm}(k)= -sh + \frac{J}{2m}k \pm \sqrt {\left (\frac{k^2}{2m}-\mu \right )^2 + \Delta^2}$, where $s=\pm$, with $+$ and $-$ corresponding to different spin configurations depending on the orientation of the $\bm{h}$.


For translationally invariant systems p-wave triplet correlations derive from equation 

\bea
\quad  <\Delta_{\bm{p}}> \propto \frac{1}{\beta}\Sigma_{i\omega_n} Tr \{k\tau_x \bm{d} \cdot \tilde{\bm{\sigma}}i\sigma_y\hat{G}(k,i\omega_n)\}\,,
\eea

\noi where $\hat{G}(k,i\omega_n)$ the Matsubara Green's function with $\omega_n=(2n+1)\pi/\beta$ the fermion frequencies and $\beta \propto 1/T$ the inverse of temperature. The above expression simplifies to 
\bea
\quad  <\Delta_{\bm{p}}> \propto \sum_{k}\sum_{m} && n_{F}(E_m) \no \\
&&[U(k)^{\dag} k\tau_x \bm{d} \cdot \tilde{\bm{\sigma}}i\sigma_y U(k)]_{mm} \,,
\eea
\noi where $m$ the energy band index, $U(k)$ the transformation matrix which diagonalises the Hamiltonian matrix $\hat{H}(k)$ while $[]_{mm}$ denotes the $m$ diagonal term of the corresponding matrix. Based on the above analysis it is straightforward to derive the expression

\bea
 &&<\Delta_{p,\bm{d} \parallel  \bm{h}}> \propto \sum_{k} k \frac{\Delta}{\sqrt{\Delta^2+(\frac{k^2}{2m}-\mu)^2}} \cdot \no \\
&&\biggl[ n_F(E_{s,-})-n_F(E_{s',+}) - [n_F(E_{s',-})  - n_F(E_{s,+}) \biggl] \label{Eq:2}
 \eea

 \noi for the triplet correlations with $\bm{d} \parallel \bm{h}$. Since for $h=0$, $n_F(E_{s,\pm})=n_F(E_{s',\pm})$, triplet correlations vanish, $<\Delta_{p,d \parallel  \bm{h}}>=0$. Similarly, for $J=0$ we get $<\Delta_{p,d \parallel  \bm{h}}>=0$, since 
 the expression $k\left (n_F(E_{s,-})+n_F(E_{s',+}) \right.$ \\
  $\left.  - [n_F(E_{s,+})  + n_F(E_{s',-})] \right )$ is odd in momenta $k$. However, when both $h$ and $J$ are finite, both spin and spatial inversion symmetries break, leading to a finite sum for triplet correlations $<\Delta_{p,d \parallel  \bm{h}}>$.  Regarding correlations $<\Delta_{p,d \perp \bm{h}}>$, matrix $[U(k)^{\dag} k\tau_x \bm{d}_{\perp \bm{h}} \cdot \tilde{\bm{\sigma}}i\sigma_y U(k)]$ has no diagonal elements, therefore they vanish. Thus, only when $h$ and $J$ are both finite in a conventional superconductor $\Delta$, $<\Delta_{p,d \parallel  \bm{h}}>$ correlations are induced. From Eq. \ref{Eq:2} it becomes apparent that $<\Delta_{p,d \parallel  \bm{h}}>$ correlations emerge from the imbalance among the two configurations $\psi_{k,s}^{\dag}\psi_{-k,s'}^{\dag}$ and $\psi_{k,s'}^{\dag}\psi_{-k,s}^{\dag}$ of the interband singlet pairing (See Sup. Mat.). Moreover, since $J \rightarrow -J$ leads to $E_{s,\pm}(k) \rightarrow E_{s,\pm}(-k)$ we get $<\Delta_{p,d \parallel  \bm{h}}> \rightarrow -<\Delta_{p,d \parallel  \bm{h}}>$ or equivalently the phase of the triplet order parameter changes by $\pi$.  Most importantly, there are no threshold values for $h$ and $J$ in order for $<\Delta_{p,\bm{d} \parallel  \bm{h}}> $ correlations to be induced. Therefore the particular mechanism is valid even for small values of these parameters which leave conventional superconductivity unaffected.  \\
\section{Topological superconductivity in superconductor-ferromagnet heterostructures}
\label{sec:2}
\begin{figure}[t!]
\includegraphics[scale=0.2]{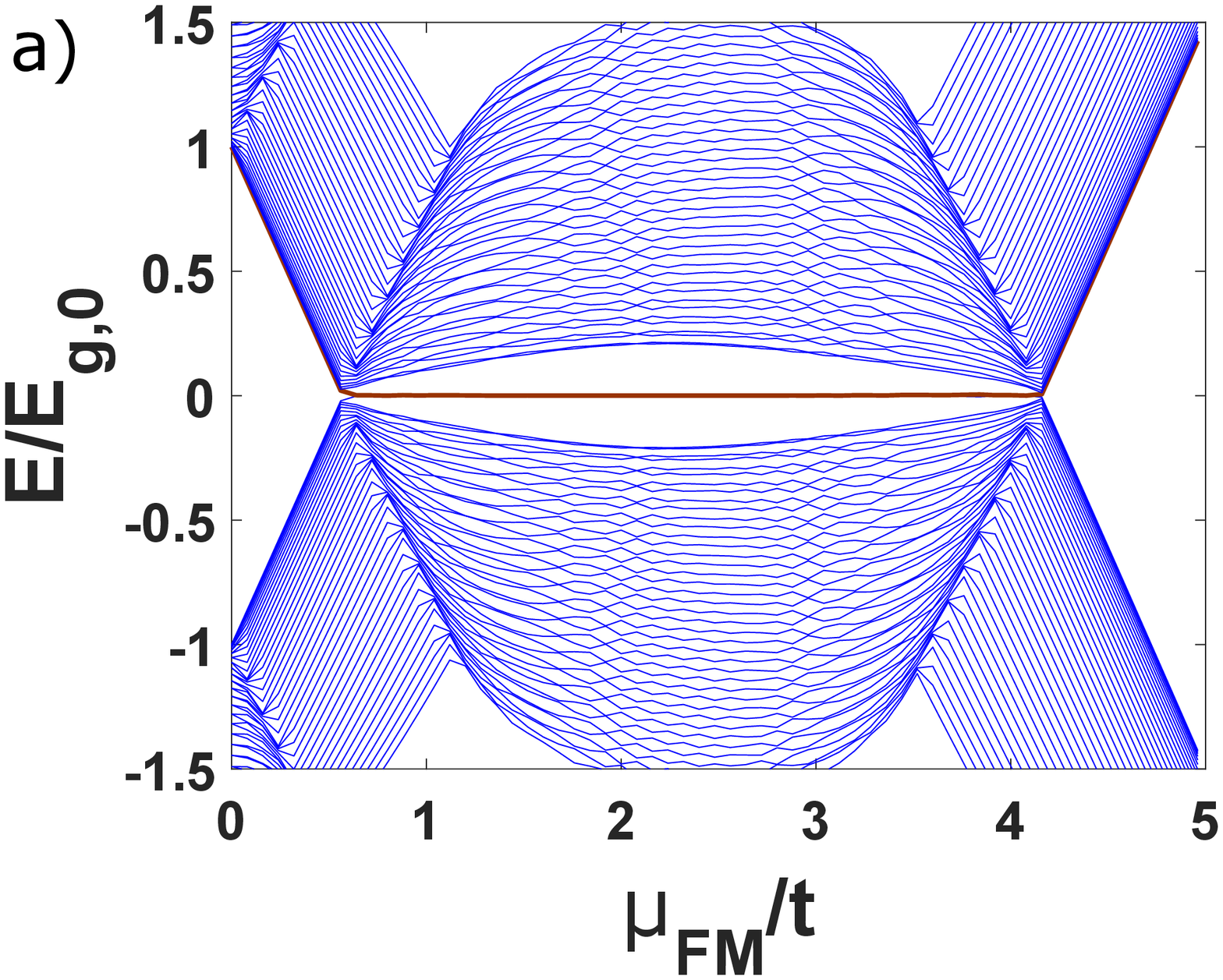}
\includegraphics[scale=0.2]{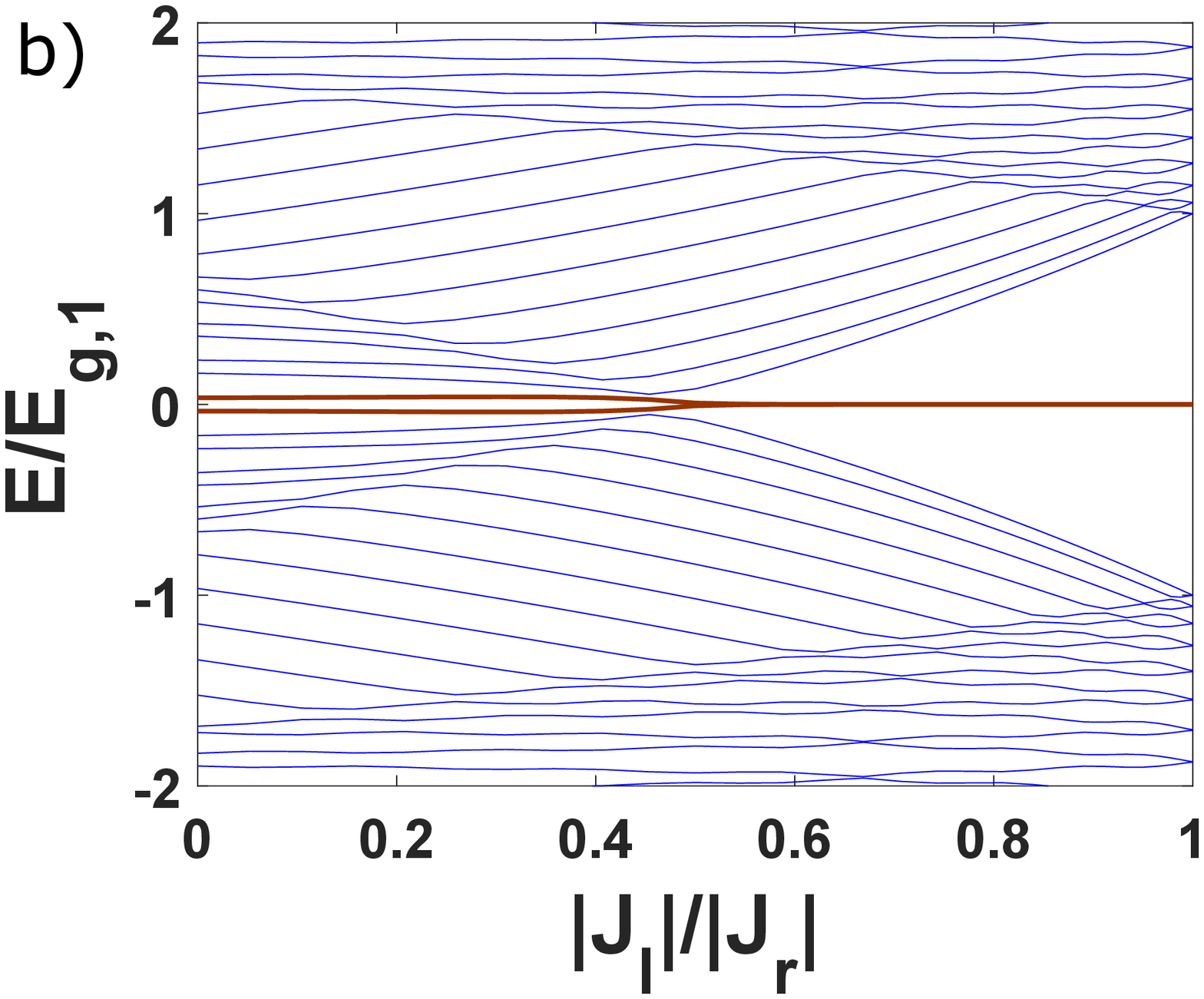}
\includegraphics[scale=0.2]{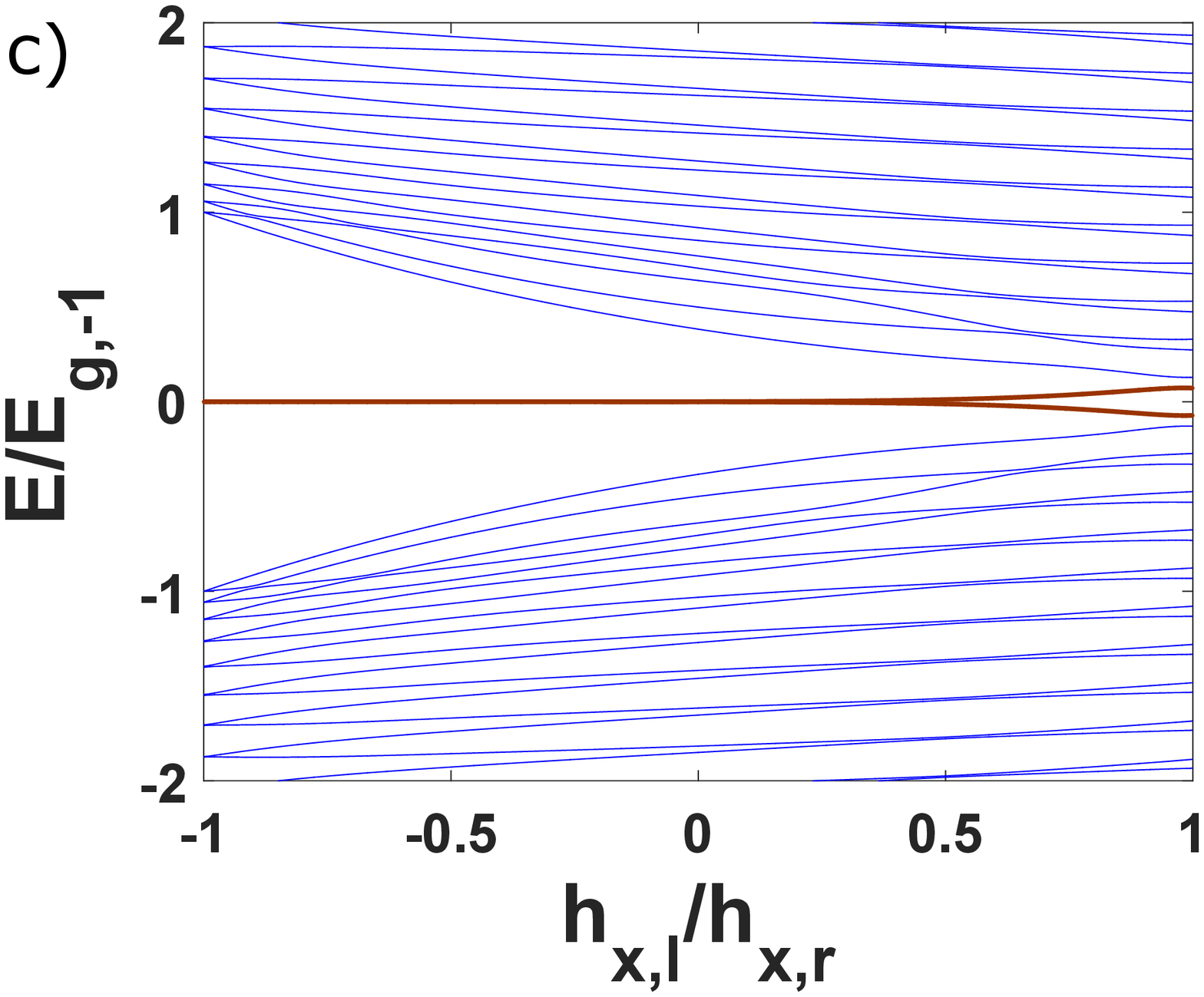}
\includegraphics[scale=0.2]{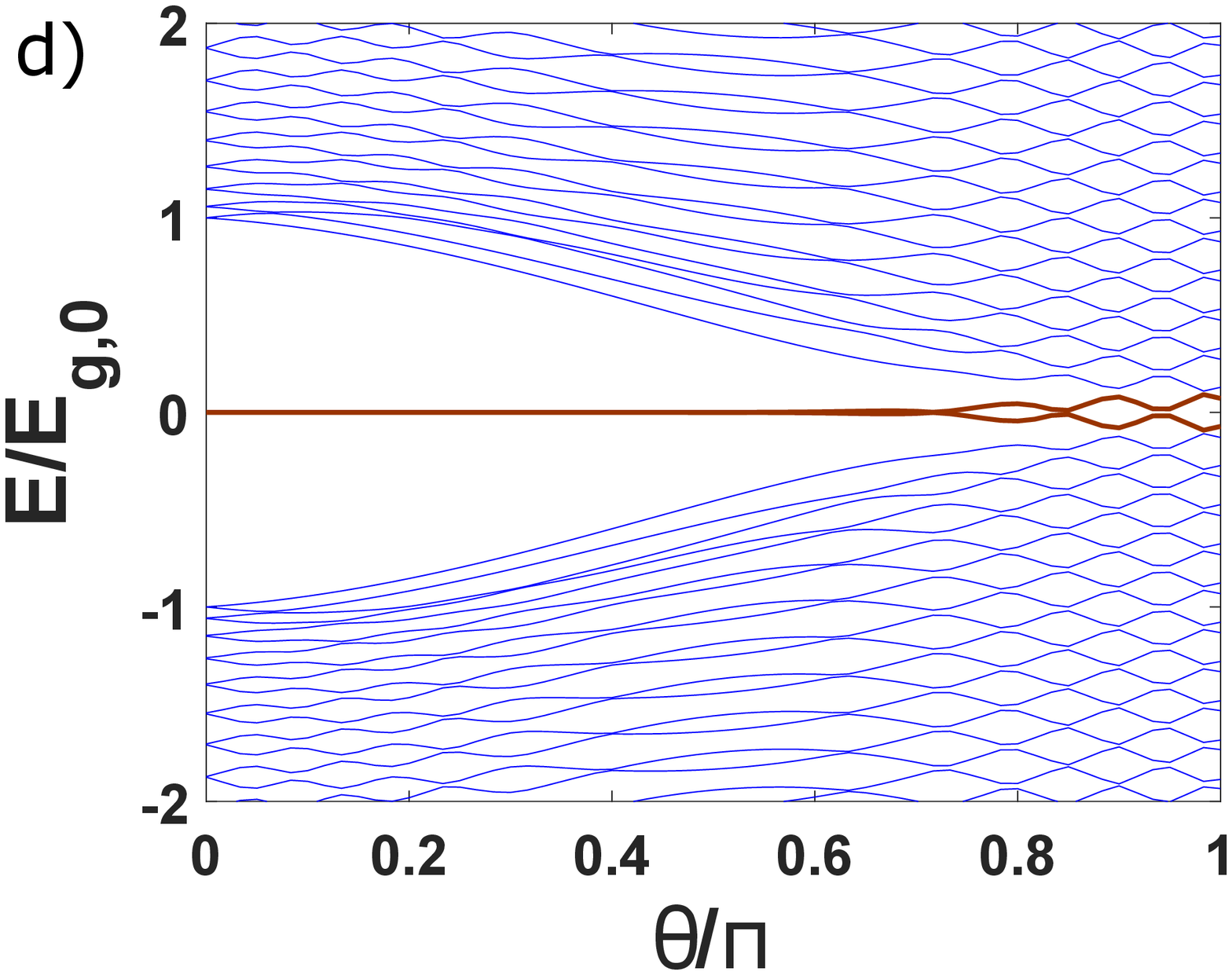}
\caption{a) The low energy spectrum of the device as a function of the chemical potential of the wire $\mu_{FM}/t$. Eigenenergies are normalised with respect to energy gap $E_{g,0}$  for $\mu_{FM}=0$ and $h_{FM}/t=2.4$. b) The energy gap in the wire for $h_{FM}/t=\mu_{FM}/t=2.4$ with respect to the ratio of the magnitude of the supercurrents $|J_{l}|$,$|J_{r}|$ in the left and the right SC, respectively. The supercurrents flow in the opposite direction. The energy spectrum is normalised to the energy gap $E_{g,1}$ corresponding to $J_{l}=-J_{r}$. c) Low energy excitation spectrum as a function of the ratio $h_{x,l}/h_{x,r}$ of the exchange fields along the $x$-axis in the left and right SC. The antiparallel configuration leads to the maximum energy gap. d) The dependence of the excitation spectrum on the rotation angle $\theta$ of $h_{r}$ over the right SC in the $x-z$ plane, where for $\theta=0$ $h_{r}=-h_{l}$ along the $x$-axis. Energies in c) are normalised to $E_{g,-1}$ corresponding to $h_{x,l}=-h_{x,r}$, while in d) are normalised to  $E_{g,0}$ for $\theta=0$. The device appears to be more robust against deviations from the optimal anti-parallel configuration of the exchange fields.}
 \label{fig:2}
\end{figure}

\indent Evidently, topological superconductivity cannot be realised simply by applying a supercurrent in a superconducting wire under the presence of an exchange field, since the emergent triplet pairing correlations are parallel to the polarization of the exchange field and therefore connect different energy bands of the wire, i.e. we get interband triplet pairing. In order to realise MZMs we need intraband superconductivity. In our device intraband superconductivity emerges in the FM where the triplet correlations induced in the conventional SCs mediate by means of the proximity effect. Necessary condition for the emergence of intraband superconductivity is that the exchange field $h$ of the wire and the $\bm{d}$ vector of the induced triplet correlations are misaligned. Therefore, a topological superconducting state is realised in a FM separating two conventional SCs when supercurrents are applied in an antiparallel configuration and the exchange fields are perpendicular to the magnetisation of the wire and antiparallel to each other (see Fig. \ref{fig:1}). The antiparallel configuration guarantees that the supercurrent and the exchange field vanish into the wire and therefore no $<\Delta_{p,d \parallel \bm{h}_{FM}}>$ correlations are induced there. Although this is the optimal configuration, we show that TSC is realised even for small misalignment angles among the magnetisations of the FM and the FIs and even for not exactly anti-parallel supercurrents and exchange fields (see Fig: \ref{fig:2}).


In order to illustrate numerically the above arguments we employ the following lattice Bogoliubov de Gennes equation, ${\cal H} = \sum_{\bm{i},\bm{j}}\Psi_{\bm{i}}^{\dag}H_{\bm{i},\bm{j}}\up\Psi_{\bm{j}}\up$, which describes SC-FM presented in Fig. \ref{fig:1}, with
\bea
\quad  H_{\bm{i},\bm{j}}= (tf_{\bm{i},\bm{j}}+\mu_{\bm{i}})\tau_z + \tau_z\bm{h_{\bm{i}}} \cdot \tilde{\bm{\sigma}} + \Delta_{\bm{i}}\tau_y\sigma_y + \bm{J_{\bm{i}}} \cdot \bm{g}_{\bm{i},\bm{j}} \label{Eq:3}
\eea

\noi where $\Psi_{\bm{i}}^{\dag} = (\psi_{\uparrow,\bm{i}}^{\dag},\psi_{\downarrow,\bm{i}}^{\dag},\psi_{\uparrow,\bm{i}}\up,\psi_{\downarrow,\bm{i}}\up)$ is the Nambu spinor, $\bm{\tau}$ and $\tilde{\bm{\sigma}}=(\sigma_x,\tau_z\sigma_y,\sigma_z)$ the Pauli matrices acting on particle-hole and spin space respectively, $f_{\bm{i},\bm{j}}=\delta_{\bm{j},\bm{i} \pm \hat{x}} + \delta_{\bm{j},\bm{i} \pm \hat{y}}$ and $\bm{g}_{\bm{i},\bm{j}}=(\pm i\delta_{\bm{j},\bm{i}\pm \hat{x}}, \pm i \delta_{\bm{j},\bm{i}\pm \hat{y}})$ the even and odd in spatial inversion, respectively, functions connecting nearest neighbours lattice points. Moreover, $t$ is the hopping integral, $\mu_{\bm{i}}$ the chemical potential, $\bm{h_{\bm{i}}}$ the exchange field, $\Delta_{\bm{i}}$ the conventional superconducting field and $J_{\bm{i}}$ the supercurrent term, on site $\bm{i}$. Hamiltonian Eq. \ref{Eq:3} describes the SC-FM-SC heterostructure considering the appropriate values of $\Delta_{\bm{i}}$, $\mu_{\bm{i}}$, $\bm{h_{\bm{i}}}$ and $J_{\bm{i}}$ for each region. The proximity to the FIs is modelled by considering a finite exchange field $\bm{h}$ over the SCs emerging from the interfacial exchange interaction between the localized magnetic moments of the FIs and the SCs conduction band electrons. MZMs are anticipated to emerge localised at the edges of the wire for $|2t - \sqrt{(h_{FM})^2-\Delta'^2}|< |\mu_{FM}| <|2t + \sqrt{(h_{FM})^2-\Delta'^2}|$ where $\Delta'$ is the induced by proximity singlet superconducting field over the FM \cite{Livanas}. From Fig. \ref{fig:2}a we conclude that the hopping term along the wire is renormalised to $t'\simeq 0.85t$, due to the coupling of the wire to the SCs \cite{Peng}. Figs. \ref{fig:2} b),c) and d) demonstrate the significant robustness of our device against deviations from the optimal antiparallel alignment of the magnetisation in the FIs  and the antiparallel configuration of the supercurrents in the SCs.

\section{Controlling Majorana zero modes}
\label{sec:3}

\begin{figure}[!ht]
\includegraphics[scale=0.8]{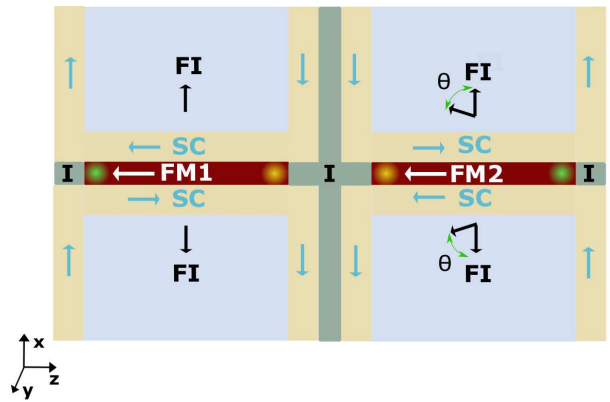}
\caption{ Setup of two ferromagnets (FMs) separated by an insulator (I) in proximity with conventional superconductors (SCs) and ferromagnetic insulators (FIs). The junction is controlled by varying the supercurrents (blue arrows) and/or rotating the magnetisation (black arrow) of the FIs adjacent to FM2 by angle $\theta$.}
\label{fig:3}
\end{figure}

\begin{figure}[!t]
  \includegraphics[scale=0.18]{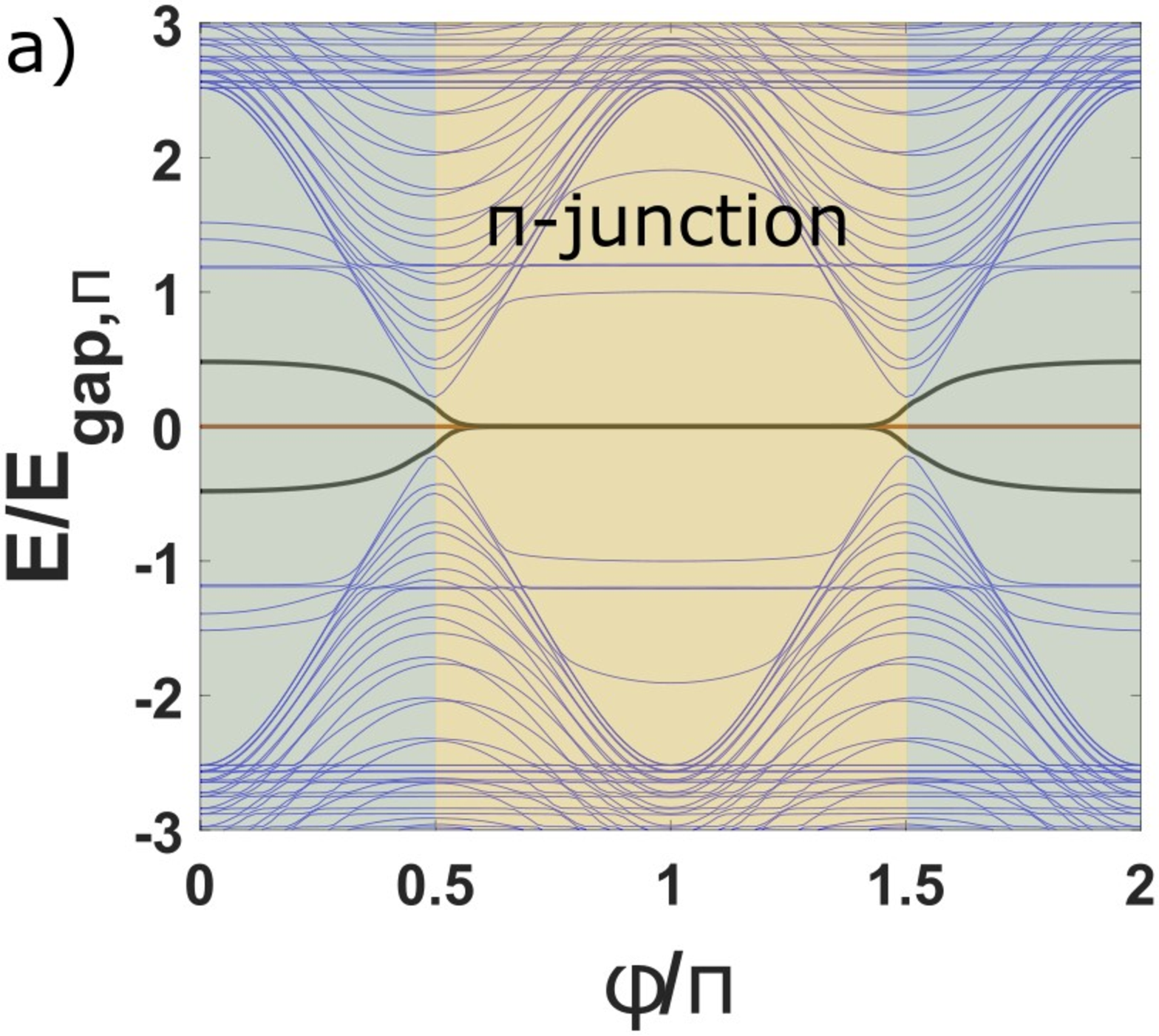}
  \includegraphics[scale=0.18]{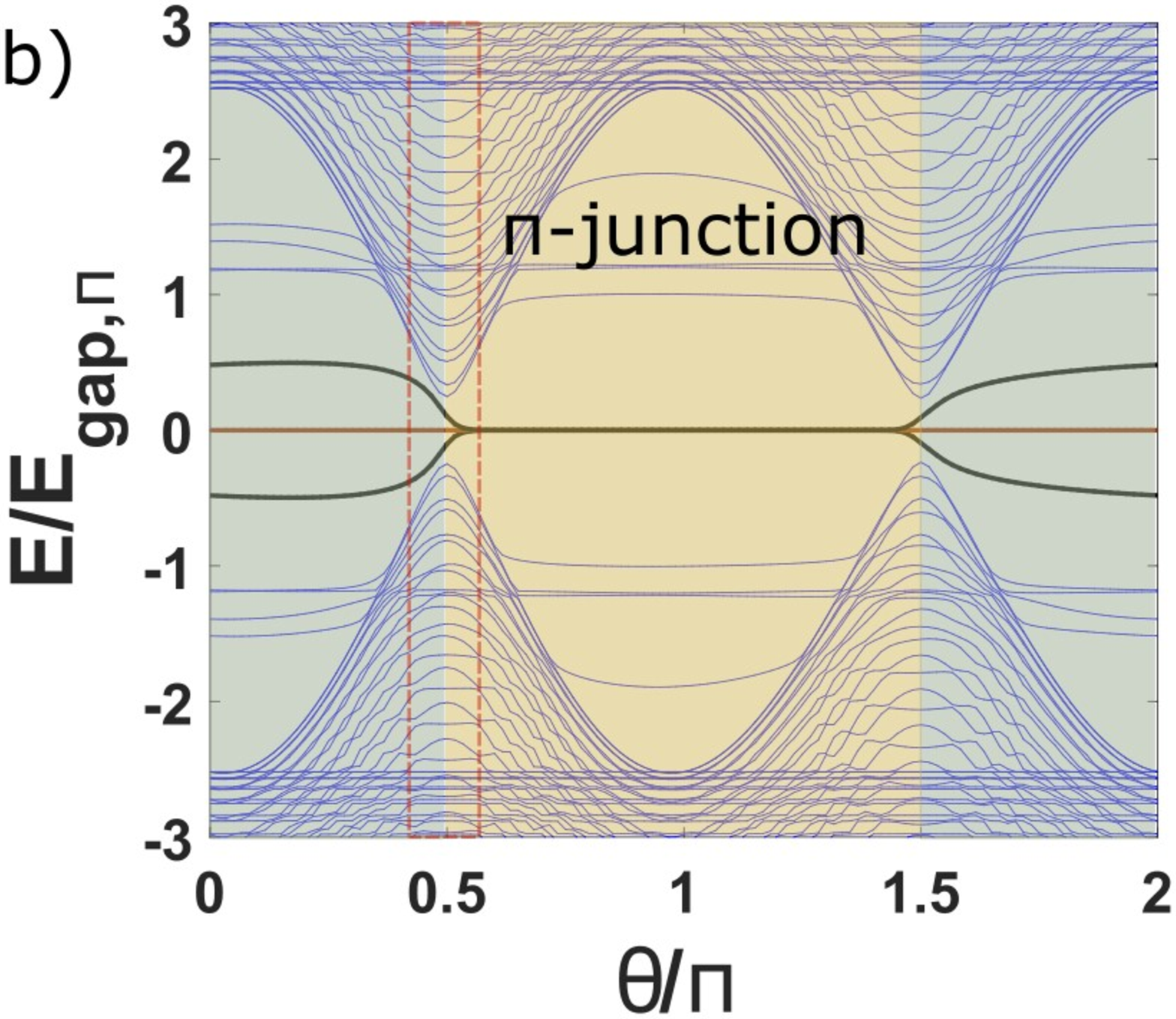}\\
  \qquad \includegraphics[scale=0.2]{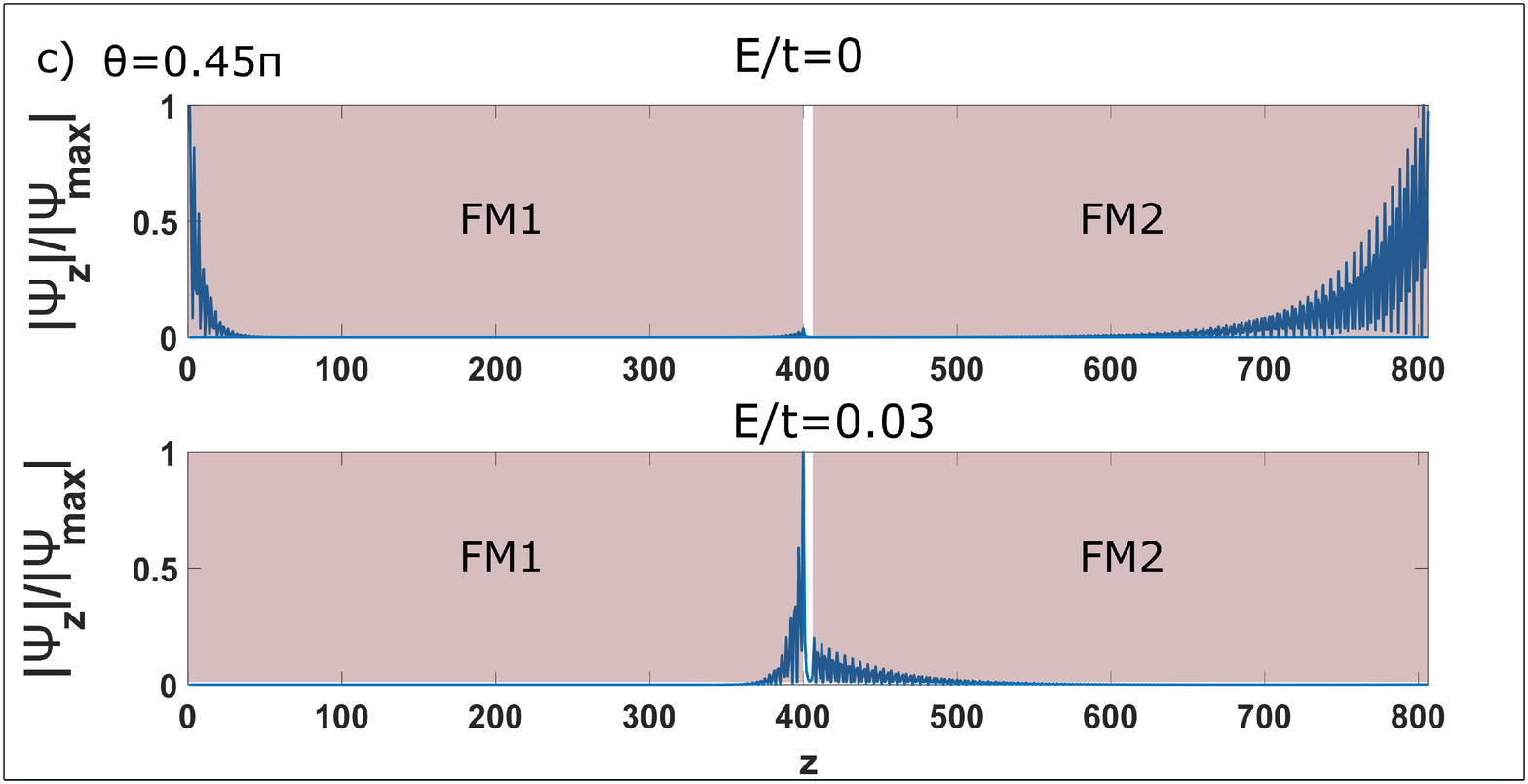}
  \qquad \includegraphics[scale=0.2]{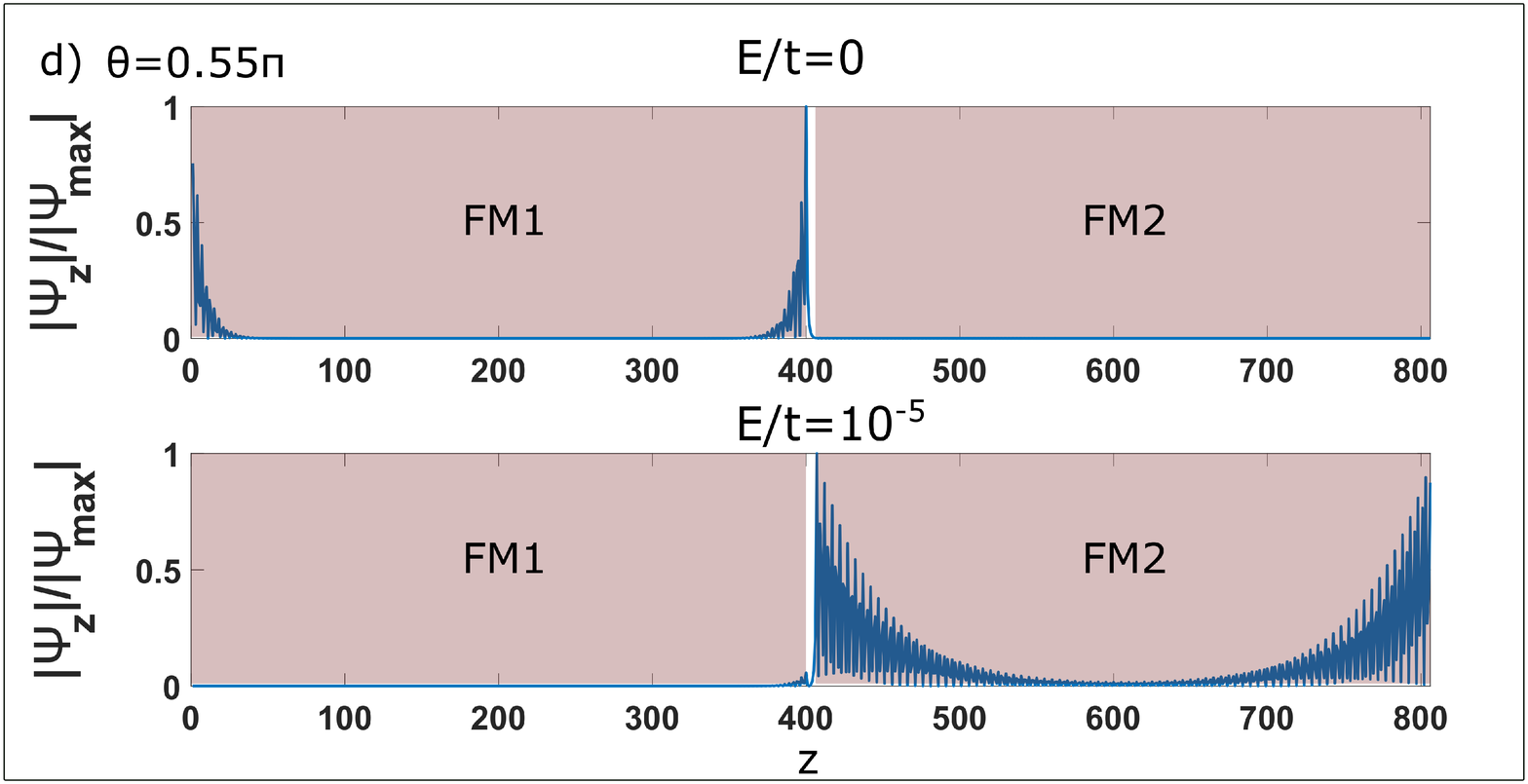}
  \caption{Low-energy quasiparticle energy spectrums for the setup presented in Fig. \ref{fig:3}, a) with respect to parameter $\phi$ controlling the supercurrents ($J=J_0\cos(\phi)$) in the conventional SCs adjacent to FM2 and b) with respect to $\theta$ corresponding to the angle of FIs magnetisation in the x-z plane. Eigenenergies are normalised to the energy gap $E_{gap,\pi}$ for $\phi=\pi$ and $\theta=\pi$ respectively. The second lowest eigenenergy (black line) is multiplied by 10 to demonstrate clearly the splitting of the MZMs. The spectral function $|\Psi_{z}|$ of the two lowest eigenenergies of the heterostructure, c) for $\theta=0.45\pi$ and d) for $\theta=0.55\pi$. The particular angles are indicated in figure b) by the dashed red line. The function is normalised to the maximum value $|\Psi_{max}|$. For $\theta=0.45\pi$, MZMs localised in the right edge of FM1 and left edge of FM2 form a fermionic state with finite energy $E/t=0.03$. For $\theta=0.55\pi$, four MZMs emerge. Since in FM2 the FIs magnetisation is not in the optimal anti-parallel configuration, MZMs appear to be less localised. In the upper c) plot the left and the right part of the function is normalised independently in order to compare the localisation length of MZMs in the two FMs.}
  \label{fig:4}
\end{figure}

To demonstrate the controllability of our setup 
we consider a junction between two ferromagnetic wires FM1 and FM2 as depicted in Fig. \ref{fig:3}. The conventional superconductors in this configuration are separated by an insulator in order for the supercurrents adjacent of FM1 and FM2 to be controlled independently. First, we consider a sinusoidal variation $J=J_0\cos(\phi)$ of the magnitude of the supercurrents applied in the SCs that are adjacent to the FM2. The supercurrents always flow in opposite directions. In Fig. \ref{fig:3} is presented the supercurrents configuration for $\phi=\pi$ while in Fig. \ref{fig:4}a) is shown the low-energy spectrum of this junction with respect to parameter $\phi$. 
The FM wires are coupled through the insulating region with a hopping term $t_c=0.5t$. For $-\pi/2<\phi<\pi/2$, the superconducting states in the two FMs acquire the same phase ($0$- Josepshon junction) and MZMs $\gamma_{1,b}$ localised in the right edge of FM1 and $\gamma_{2,a}$ localised at the left edge of the FM2  interact with each other giving rise to an ordinary fermionic state with finite energy. The other two MZMs $\gamma_{1,a}$ at the left edge of FM1 and $\gamma_{2,b}$ at the right edge of FM2 remain unpaired and therefore stick at zero energy. For $\phi= \pm \pi/2$ the energy gap in FM2 closes and MZMs $\gamma_{2,a}$ and $\gamma_{2,b}$ fuse while $\gamma_{1,a}$ and $\gamma_{1,b}$ remain uncoupled. Remarkably for $\pi/2<\phi<3\pi/2$, for which the supercurrent flow is reversed on both sides of FM2, \emph{the phase of the induced p-wave superconductivity in FM2 changes by $\pi$} leading to the formation of a \emph{$\pi$-Josepshon junction} among the FM wires. Since the reversal of the flow of the supercurrent leads to an \emph{interchange of the position of $\gamma_{2,a}$ and $\gamma_{2,b}$}, and $\gamma_{2,b}$ cannot couple with $\gamma_{1,b}$, the heterostructure hosts four MZMs reflected in the low-energy quasi-particle spectrum. Note on the same figure that the four MZMs are stabilised even for very small values of the supercurrent.

Next we investigate the same junction by rotating the magnetisation and therefore the exchange field of the FIs adjacent to FM2, in the x-z plane, $h=h_z\sin(\theta)+h_x\cos(\theta)$ where $\theta$ is the angle from the x-axis. Since we require that the $x$ component of the exchange fields of FIs adjacent to FM2, point at the opposite direction for every $\theta$, the magnetisation 
in the upper FI  rotates clockwise while in the lower  FI 
 rotates counter-clockwise (Fig.\ref{fig:3}).  The supercurrent in the two FMs flows along the same direction, therefore for $\theta=0$ the junction is formed by two identical copies of FMs embedded in conventional SCs. According to our mechanism topological SC emerges only when the magnetisation of the FIs and that of the ferromagnetic wire is misaligned. Indeed, for $\theta=\pi/2$ the energy gap at the wire closes (Fig. \ref{fig:4}b)). Remarkably for $\pi/2<\theta<3\pi/2$ for which the $x$ component  of the magnetisation of the FIs adjacent to FM2 is opposite to that of the FIs adjacent to FM1, \emph{a $\pi$-Josepshon junction is formed} and four Majorana zero modes emerge localised at the edges of the two wires. In Fig. \ref{fig:4}c) we display the spectral function $|\Psi_z|$ of the two lowest eigenvalues for $\theta=0.45\pi$ and $\theta=0.55\pi$. We remark that small rotations of the FIs magnetisation can be achieved by the application of gate voltages. 


\section{Conclusions}
\label{sec:4} We present a novel device platform for engineering MZMs  based on SC-FM heterostructures. Key elements of the underlying mechanism are the externally applied supercurrents and the emergent exchange fields, due to proximity of the SCs with FIs. Developments in the fabrication of SC-FI heterostructures \cite{Liu} and ferromagnetic metallic Co \cite{Giroud,Wang} or Fe \cite{Yazdani} nanonwires embedded in conventional SCs, enhance the experimental feasibility of our proposal. Several works have demonstrated both theoretically \cite{Tokuyasu,Hijano} and experimentally \cite{Tedrow,Hao,Moodera} the emergence of exchange fields in conventional SCs in proximity to FIs, e.g. in EuS/Al and EuO/Al heterostructures. Finally, our platform does not require externally applied magnetic fields, yet it facilitates the control of the emergent MZMs, either through the applied supercurrents or the manipulation of FIs magnetisation  by use of currents or applied gate voltages. In this way we circumvent the screening effects problem that undermines schemes for braiding MZMs based on semiconductors.

This research is carried out in the context of the project "Majorana fermions from induced quartet states" (MIS 5049433) under the call for proposals "Researchers' support with an emphasis on young researchers- 2nd Cycle". The project is co-financed by Greece and the European Union (European Social Fund- ESF) by the Operational Programme Human Resources Development, Education and Lifelong Learning 2014-2020.”

\paragraph{Author contribution statement}

Dr. Giorgos Livanas has conducted the analysis and wrote the initial version of the manuscript, Nikos Vanas contributed to numerical calculations, Prof. Giorgos Varelogiannis and Prof. Manfred Sigrist have revised and wrote the final version of the manuscript. The work was conceptualized by Dr. Giorgos Livanas and Prof. Giorgos Varelogiannis.   




\end{document}